\documentclass[runningheads]{llncs}
\usepackage{graphicx}
\usepackage{graphicx}
\usepackage{hyperref}
\usepackage{ifthen}
\usepackage{textcomp}
\usepackage{xcolor}
\usepackage{svg}
\usepackage{cite}
\usepackage{amsmath,amssymb,amsfonts}
\usepackage{algorithmic}
\usepackage{graphicx}
\usepackage{caption}
\usepackage{subcaption}
\usepackage{fancyvrb}
\usepackage{xcolor}
\usepackage{listings}
\usepackage{xcolor} 

\usepackage[T1]{fontenc}
\newboolean{showcomments}
\setboolean{showcomments}{true} 
\ifthenelse{\boolean{showcomments}}
  {\newcommand{\nb}[2]{
    \fcolorbox{gray}{yellow}{\bfseries\sffamily\scriptsize#1}
    {\sf\small$\blacktriangleright$\textit{#2}$\blacktriangleleft$}
   }
   
  }
  {\newcommand{\nb}[2]{}
   
  }

\definecolor{codered}{RGB}{177,84,105}
\definecolor{codeblue}{RGB}{86,53,255}
\definecolor{codegreen}{RGB}{63,127,95}
\definecolor{codesalmon}{RGB}{250,128,114}

\begin{document}
\title{Vibe Modeling: Challenges and Opportunities}
\titlerunning{Vibe Modeling: Challenges and Opportunities}
%

\author{Jordi Cabot\inst{1,2}}

%
\authorrunning{J. Cabot}

%
\institute{Luxembourg Institute of Science and Technology, Esch-sur-Alzette, Luxembourg \\
\email{jordi.cabot@list.lu} \and
University of Luxembourg, Esch-sur-Alzette, Luxembourg }
\maketitle              
\begin{abstract}

There is a pressing need for better development methods and tools to keep up with the growing demand and increasing complexity of new software systems. 
New types of user interfaces, the need for intelligent components,  sustainability concerns, ... bring new challenges that we need to handle. 

In the last years, model-driven engineering (MDE) has been key to improving the quality and productivity of software development, but models themselves are becoming increasingly complex to specify and manage. 
At the same time, we are witnessing the growing popularity of vibe coding approaches that rely on Large Language Models (LLMs) to transform natural language descriptions into running code at the expenses of code vulnerabilities, scalability issues and maintainability concerns.

In this paper, we introduce the concept of \textit{vibe modeling} as a novel approach to integrate the best of both worlds (AI and MDE) to speed up the development of reliable complex systems.
We outline the key concepts of vibe modeling and highlight the opportunities and open challenges it presents for the future of modeling.

\keywords{Vibe Modeling \and Low-modeling \and Low-code \and DSL \and Artificial Intelligence \and Model-driven \and Vibe Coding.}
\end{abstract}

\section{Introduction}

Current software development projects face a growing demand for advanced features, including support for new types of user interfaces (augmented reality, virtual reality, chat and voice interfaces,...), intelligent behavior to be able to classify/predict/recommend information based on user's input or the need to face new security and sustainability concerns, among many other new types of requirements.   

To tame this complexity, software engineers typically choose to work at a higher abstraction level~\cite{Booch18} where technical details can be ignored, at least during the initial development phases.
{\em Low-code platforms} are the latest incarnation of this trend, promising to accelerate software delivery by dramatically reducing the amount of hand-coding required. 
Low-code can be seen as a continuation or specific style of other model-based approaches~\cite{cabot2024lowcode, RuscioKLPTW22}, where high-level software models are used to (semi)automatically generate the running software system. 

However, even models themselves are becoming more and more complex due to the increasing complexity of the underlying systems being modeled. 
Beyond ``classical'' data and behavioral aspects, we now need to come up with new models to define the new types of UIs or all the smart features of the system. 
Note that AI elements are hard to specify~\cite{rahimi2019toward}, architect, test and verify~\cite{RiccioJSHWT20} and low-code systems have so far paid little attention to the modeling and development of smart systems. 

This hampers the adoption of model-driven processes as it reduces the (perceived?) return on investment of modeling activities due to the increase cost of modeling\footnote{Note that the adoption of modeling practices is a complex sociotechnical problem \cite{HutchinsonWR14}.}. 
Even more at a time when a growing number of tools offer a vibe coding approach to generate code from natural language descriptions. 
Vibe coding is an approach to producing software by using Large Language Models (LLMs), where a person describes a problem in a few natural language sentences as a prompt that is then sent to an LLM tuned for coding. 
The LLM generates software based on the description, shifting the programmer's role from manual coding to guiding, testing, and refining the AI-generated source code \footnote{\url{https://en.wikipedia.org/wiki/Vibe_coding}}. 
With the new agentic capabilities provided by many IDEs and agent-based systems built on top of those LLMs, even the testing and verification of the generated code is becoming easier, where the agent itself creates tests, runs them and refines the code if it detects any issues.
While the generated code is not always correct, the quickness of the process makes it ideal for prototyping,  exploratory development and personal projects.

These were some scenarios where, until now, we were claiming low-code and model-driven approaches had a great product-market fit due to the possibility of having an automatic code generation phase. However, now vibe coding appears as a solid competitor here. While the final quality is not the same and vibe coding introduces plenty of safety, scalability and non-determinism issues, it is clear that the vibe coding approach is here to stay and will continue to grow. Indeed, this type of "vibing" approach resonates with plenty of developers and even with non-technical people who see a way to create their own apps.

In this paper, we explore whether we could combine both trends. To do so, we introduce the concept of \textit{vibe modeling} as a novel approach to integrate the best of both worlds (AI/LLMs and MDE) to speed up the development of reliable complex systems.
With vibe modeling, you can still start from a natural language description that an agent will help to transform into a set of models as part of a conversation between you and the agent. While at the same time, you will later use deterministic and certified code-generation techniques to generate the final software system from the models safely.

The next sections are organized as follows. Section~\ref{sec:state-of-the-art} reviews the state of the art in applying LLMs and AI to modeling activities. 
Section~\ref{sec:method} discusses the concept of vibe modeling and presents a vibe-modeling-centered development process, including its integration with low-code approaches. 
Next, Section \ref{sec:collaborative} extends the core concept to cover collaborative scenarios, while Section \ref{sec:tool} comments on the infrastructure required to support vibe modeling. 
Finally, we discuss open challenges and future directions before concluding the paper.

\section{State of the art}
\label{sec:state-of-the-art}

The modeling community has embraced with real interest the idea of using LLMs to assist in modeling activities \cite{BurguenoCWZ22, Lola24, RoccoRSNR25} as part of ongoing efforts to reduce the cost of modeling itself and improve its Return on Investment \cite{lowmodeling}.

Many of these works focus on inferring a (partial) model from a natural language description. 
For instance, research conducted in~\cite{CamaraTBV23},~\cite{fill_conceptual_2023}, and~\cite{chaaben_towards_2023} evaluated the potential to create domain models from textual descriptions using prompting.  
Camara et al. used zero-shot prompting to create UML class diagrams with few syntactic errors, however, the worst results were found when the model required abstractions, such as using inheritance instead of attributes or creating association classes. 
Fill et al. used GPT-4 to create domain models for Entity Relationship diagrams for conceptual modeling, BPMN diagrams for business processes, and Heraklit models for embedded systems by providing one example of the desired output\cite{fill_conceptual_2023}. 
Chaaben et al \cite{chaaben_towards_2023} experimented with the Few-shot technique using between 2 and 4 examples for the recommendation of concepts for different domain contexts, and to assist with static and dynamic domain modeling.
Other approaches tried to go beyond simple prompting techniques and experiment with chain-of-thought \cite{chen2023automateddomainmodeling} and tree-of-thought \cite{Silva24} prompting techniques for better accuracy.

Despite advances, LLM-based domain modeling solutions still face noticeable limitations. For example, in UML class diagram modeling, accurately identifying relationships among classes remains challenging~\cite{chen2024model,Silva24,wang2024llms}.

In line with this, research such as~\cite{kourani_2024} proposed a human-in-the-loop (HITL) approach to LLM-based modeling, aiming to exploit user feedback to refine and eventually enhance the quality of domain models created by LLMs. 
Moreover, this work was focused on process models and assumes the user is a modeling expert.

In this paper, we aim to generalize and contextualize these partial approaches in a fully interactive model-driven workflow, leveraging current agentic capabilities to better integrate AI advances in modeling before generating the code with rule-based deterministic code-generators. 
To the best of our knowledge, ours is the first work that focuses on exploring this integration. 

\section{A vibe-modeling-centered development process}
\label{sec:method}

We define \textit{vibe modeling} as the process of building software through conversational interaction with an LLM tuned for modeling, not coding. 
Once the models are created, a standard model-based / low-code approach can be used to generate deterministic code from those “vibed models”.  Think of vibe modeling as a model-driven vibe coding approach.

Indeed, in vibe modeling, the LLM does not aim to generate code, but rather to produce models. Then, the model-to-code step is performed with “classical” rule-based code-generation templates (or any other type of precise and semantically equivalent executable modeling \cite{mellor2002executable} techniques).

As such, vibe modeling has two major advantages over vibe coding:

\begin{itemize}
\item{\textbf{Understandable output}. A user is able to validate the quality of the LLM output (the models), even  without coding expertise. 
Models are more abstract and closer to the domain and, therefore, a user should be able to understand them with limited effort. 
True, some basic modeling knowledge may still be required, but for sure it’s much easier to validate a model than a list of lines of code.}
\item{\textbf{Reliable code-generation}. The generation process is deterministic. If the model is good, we know the code will be good (assuming a certain level of quality in the code-generation templates but that should only be verified once and for all) and there is no need to check it every time we regenerate the system, saving plenty of time.}
\end{itemize}

Combining the two, we see that, in contrast to vibe coding, vibe modeling can be useful to both technical and non-technical experts. 
Even the latter will have better chances to generate a reliable software by validating the models while, in a vibe coding approach, due to their lack of coding expertise, there would have no other option but to blindly trust the LLM.

\begin{figure*}
  \centering
  \includegraphics[width=\textwidth]{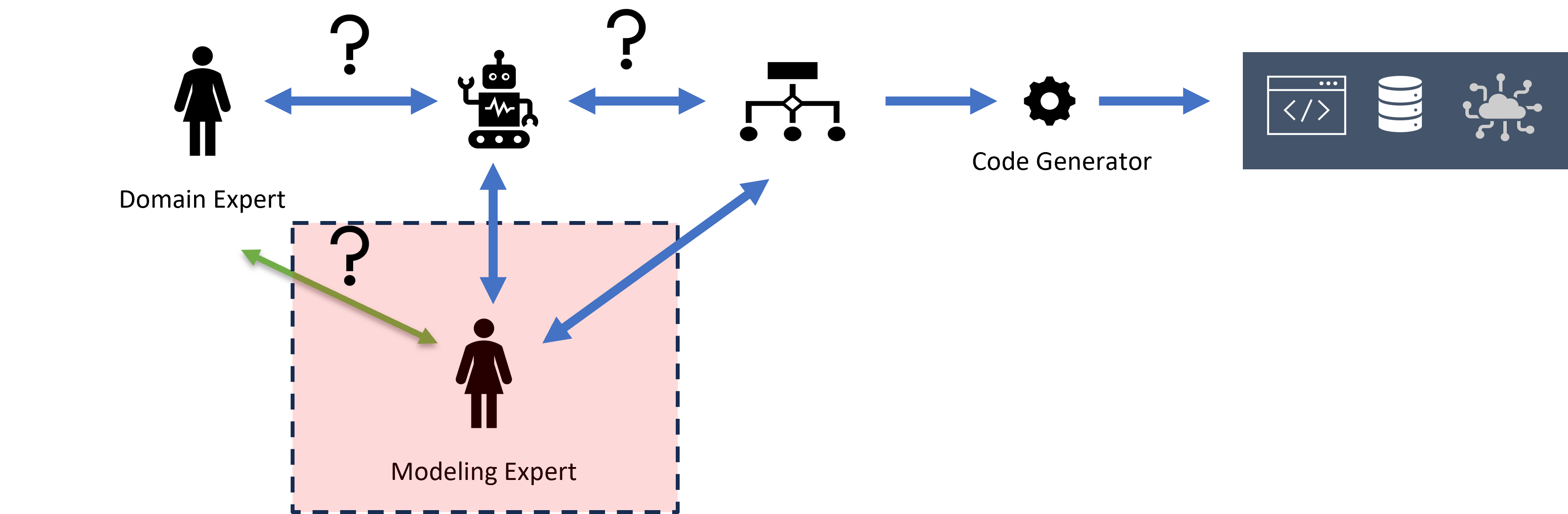}
  \caption{The vibe modeling process as part of a low-code architecture. The question mark shows interactions that carry uncertainty}
  \label{fig:arch}
\end{figure*}

Figure \ref{fig:arch} illustrates the process in more detail. 
The user (the domain expert) starts by describing the system to the agent (internally relying on a LLM). Based on this conversation (and any other material provided by the user such as interviews, guidelines,...) the agent will propose a model.
The model itself is uncertain, so the agent and the user can start a conversation to clarify and refine it. 
If a modeling expert is available and the complexity or criticality of the domain is high, the modeler can either discuss with the user to clarify ambiguities or incompleteness aspects of the description (that could then be used by the agent to improve the model) or directly refine the model.
In both cases, the modeler's role is to reduce the uncertainty on the quality of the model. 
The iteration continues until the model is considered good enough. 
At that point, the deterministic part of the process triggers in and the code-generation templates produce the final software system, including the code itself, the database (if needed), the deployment scripts (if needed), etc.

\section{Collaborative vibe modeling}
\label{sec:collaborative}

A logical evolution of our proposal is vibe modeling with, not one, but a full community of agents collaborating in real-time as the final result outperforms single-agent solutions \cite{guo2024large}.
This is starting to be explored for a variety of software engineering tasks \cite{multiSE}. 
In the specific case of modeling, this would imply to work with a community of modeling agents, each one potentially specialized in a different domain, type of model, modeling phase, etc.
Non-functional factors (like the cost, availability, sustainability or linguistic capabilities) of the agents should also play a role in the selection of the agents to be used in the collaborative modeling process.

How many agents should be involved, depending on the complexity of the model to be inferred, how much should they be specialized and whether we should put in place a purely collaborative or also a competitive approach are still open questions.
Ideally, and assuming they can afford the cost and accept the sustainability concerns, the more the merrier as the diversity of solutions proposed by the agents should lead to a better final model. 
Indeed, a purely collaborative approach has too many single points of failure, as the success of each modeling task relies on the output of a single agent. 
Instead, a competitive approach, where more than one agent is assigned the same task, is more robust but it requires putting in place a mechanism to select the best solution from the different proposals.

\begin{figure*}
  \centering
  \includegraphics[width=\textwidth]{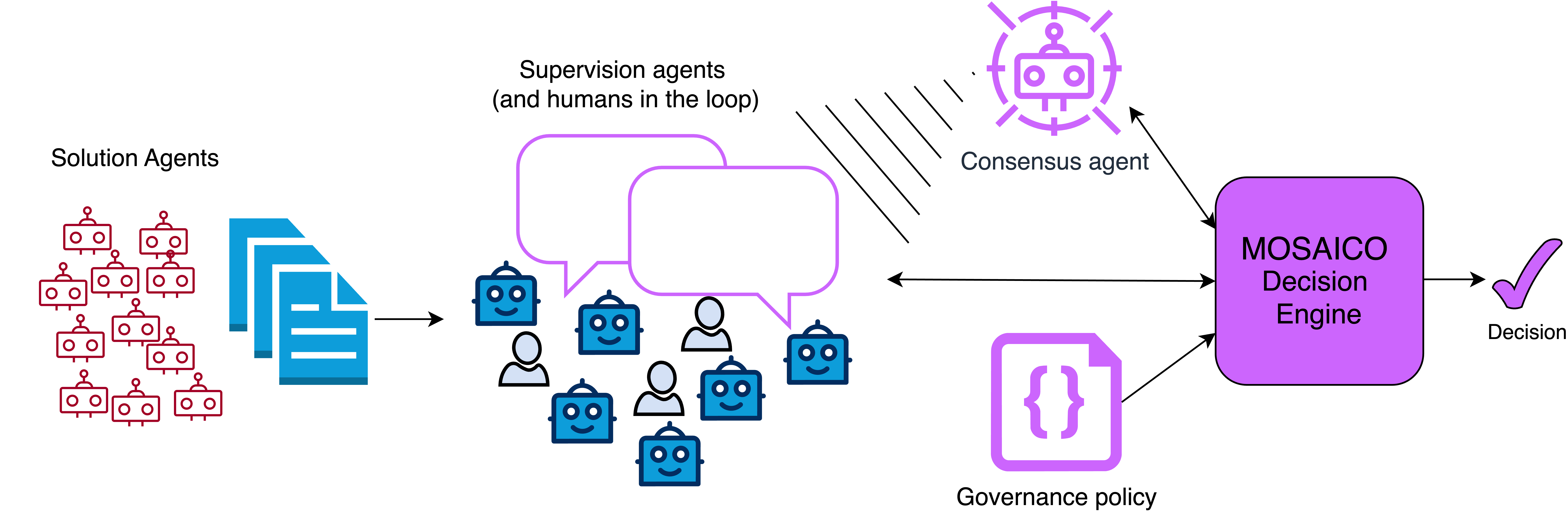}
  \caption{Overview of the AI Agent community in the Mosaico EU project}
  \label{fig:mosaico}
\end{figure*}

This "mechanism" could simply be a human in the loop. But this would require too much work from the user and may not even be feasible when the user is not a technical expert able to judge by himself the quality of the solutions.
An infrastructure like the one proposed in the Mosaico EU project \footnote{\url{https://mosaico-project.eu/}}, involving different types of agents (solution, supervision and consensus) could help here. 
In this example (see Figure \ref{fig:mosaico}) the solution agents are responsible for generating a (part of the) model, the supervision agents would then score each proposal stating which one they think it is best while the consensus agent would use these evaluations to choose the final solution. 
This choice could be based on a governance policy defined at the beginning of the modeling project, where we could state whether the consensus agent should simply take the one with more votes in favor or try to force more of a common agreement between the supervision agents. 

\section{Vibe modeling infrastructure}
\label{sec:tool}

There is a key infrastructure element to enable vibe modeling: an easy way for agents to create and manipulate models. 
We can (and should) train specialized agents to become great modelers but the agents themselves should not embed the modeling stack. 
Same as human modelers. We do not have a modeling tech stack, such as the Eclipse Modeling Framework \footnote{\url{https://www.eclipse.org/modeling/emf/}}, inside us. 
Instead, we work with Eclipse tools that expose core modeling services through different interfaces built on top of the core Eclipse components when we need to create models, view them, generate code from them, etc.

The same applies to agents. We do not want to reimplement the modeling stack as part of each agent code. Agents should be able to communicate with the modeling platform/s we want to use in our development project and benefit from the tool capabilities 
(e.g. to perform model validation, model rendering and many other basic model manipulation operations that are common to most modeling scenarios). 
But implementing a direct bridge between each agent and each modeling platform quickly triggers the MxN integration problem 
\footnote{The MxN integration problem refers to the challenge of connecting M different AI applications to N different external tools without a standardized approach \url{https://huggingface.co/learn/mcp-course/en/unit1/key-concepts}. 
It is a recurrent problem in information systems development that also appears in the context of vibe modeling.}.

\begin{figure*}
  \centering
  \includegraphics[width=0.7\textwidth]{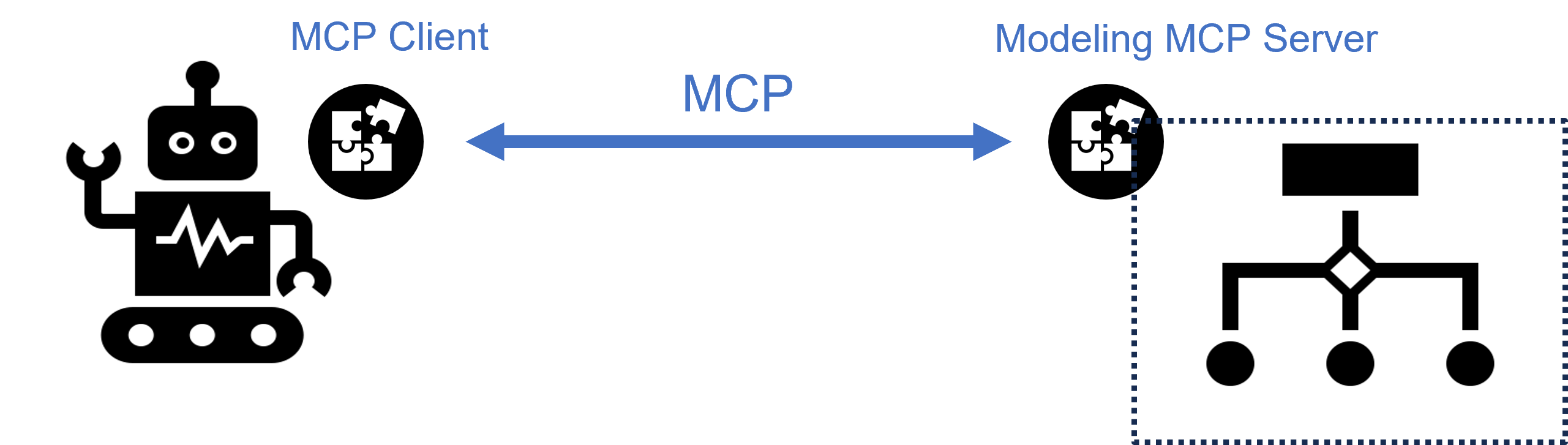}
  \caption{MCP for modeling interactions}
  \label{fig:mcp}
\end{figure*}

\begin{figure*}
  \centering
  \includegraphics[width=0.7\textwidth]{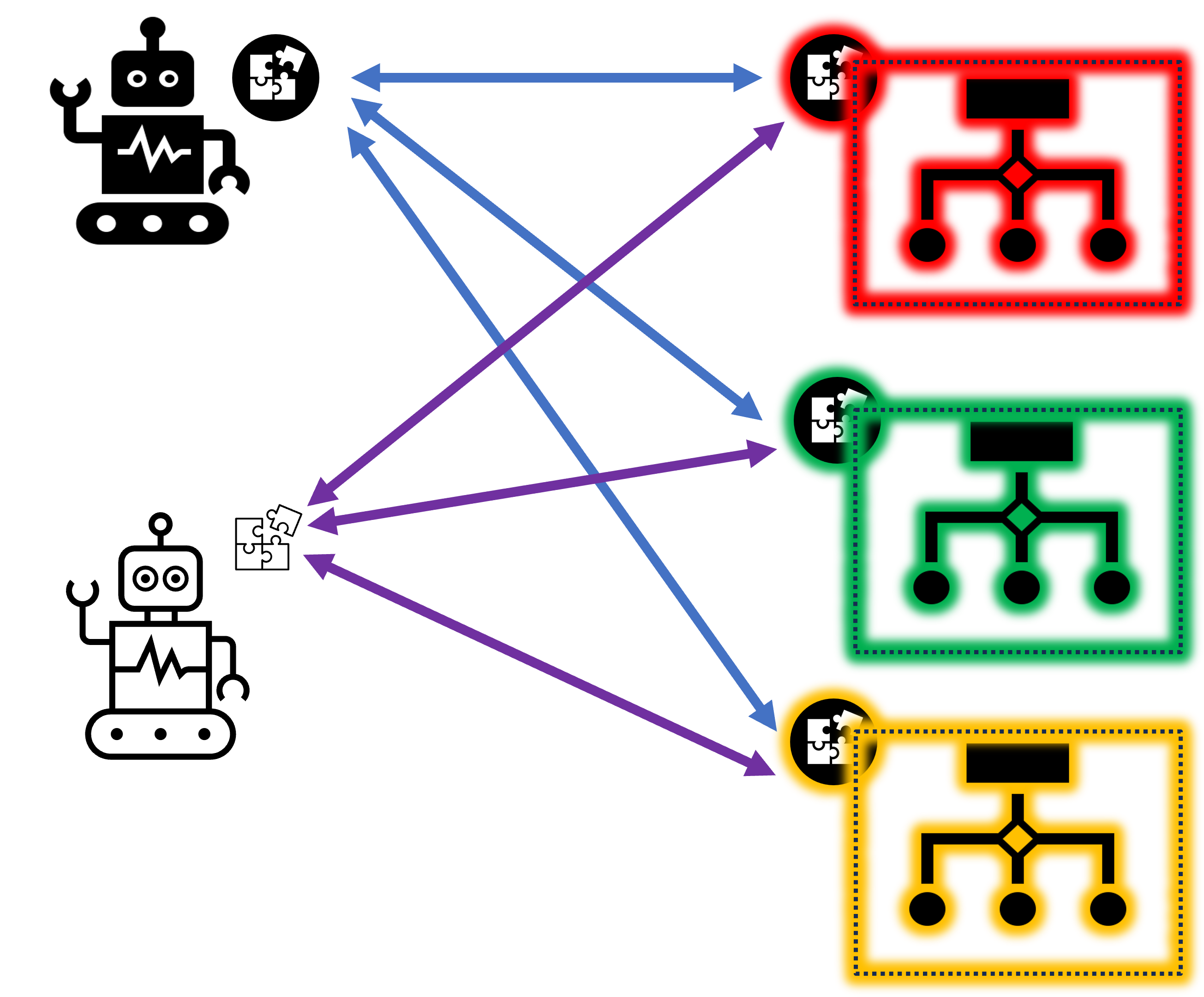}
  \caption{Multi-agent and multi-tool collaboration facilitated by the use of MCP}
  \label{fig:mcp-collab}
\end{figure*}

The Model Context Protocol (MCP) \footnote{\url{https://modelcontextprotocol.io/}} is a popular open protocol that standardizes how applications provide context to agents. 
Therefore, We propose to use MCP to bridge the gap between our modeling agents and the modeling platforms. As seen in Figure \ref{fig:mcp}, the agent embeds an MCP client able to communicate with any MCP server, in particular, the one implemented by the modeling platform, 
exposing the modeling services to the agent. Once a modeling platform offers an MCP server, any agent can use it to chat with it, and the platform does not need to adapt to the type of agent or the LLM used by such agent. 
Similarly, an agent embedding an MCP client can automatically discover and use any MCP server tools available in the environment without having 
to learn and implement code to interact with the internal modeling platform API (see Figure \ref{fig:mcp-collab}). This allows for scenarios where agents could even use, at every step of the collaboration, a different modeling platform specialized on the type of modeling request they are working on.

\begin{figure*}
  \centering
  \includegraphics[width=\textwidth]{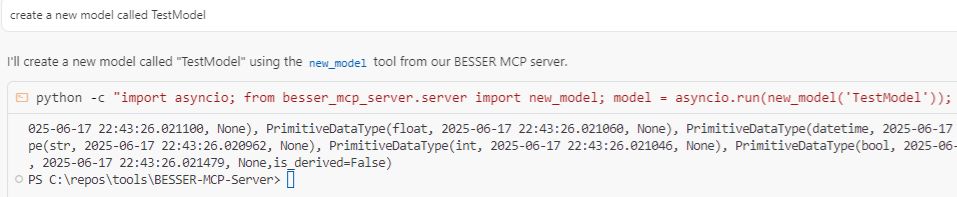}
  \caption{Example of creating a new model via a service exposed by the BESSER MCP Server}
  \label{fig:mcp-exec}
\end{figure*}

As a proof of concept, we are implementing an MCP Server for the BESSER low-code platform \cite{BESSER}. 
Thanks to this MCP server, any agent, the agent mode of Cursor\footnote{Agent is the default and most autonomous mode in Cursor, designed to handle complex coding tasks with minimal guidance \url{https://docs.cursor.com/chat/agent}} in this case, can discover and use the modeling services offered by BESSER when a user (or any other agent) requests a task for which one of the BESSER services exposed in its MCP server would be a good fit (see Figure \ref{fig:mcp-exec}).

The actual MCP Server implementation is mostly a thin wrapper on top of the internal tool API where MCP standardizes the way the tool (in a MCP context, \textit{tool} refers to a service the agent can use as a tool to achieve something, so each modeling service would be exposed as an MCP tool)
is described (and later discovered and called) by the agent. Listing~\ref{lst:besser-mcp-example} shows a simple example of the MCP Server of BESSER exposing the creation of a new B-UML model. 
Note that the model is returned serialized. This enables the agent to keep and use the model in a future interaction if needed.
An obvious alternative would be to store the model in a database and return the id of the model. 
Each approach has different trade-offs (e.g., the need to configure a database and make sure the agent has access to it) and, thus, the best approach depends on the modeling scenario.

\begin{lstlisting}[language={Python}, caption={BESSER MCP Server example}, label={lst:besser-mcp-example}]
@mcp.tool()
async def new_model(name: str) -> str:
    """Creates a new B-UML DomainModel with the specified name and returns it as base64.

    Args:
        name (str): Name of the new domain model.

    Returns:
        str: A new domain model instance as base64 string.
    """
    try:
        from besser.BUML.metamodel.structural import DomainModel  # type: ignore
    except ImportError as exc:
        raise RuntimeError(
            "BESSER library must be installed (`pip install besser`)."
        ) from exc

    # Create and return a new DomainModel instance as base64
    domain_model = DomainModel(name=name)
    return serialize_domain_model(domain_model)
\end{lstlisting}

\section{Roadmap}
\label{sec:roadmap}
So far, we have presented the key definitions and infrastructure to put in place a vibe modeling approach. 
Nevertheless, there are still a number of challenges to address to make this approach more effective and more widely adopted. 
In what follows, we discuss some open challenges. 

\subsection{Specialized modeling agents}
While there are a number of works on inferring models from natural language descriptions (see Section \ref{sec:state-of-the-art}), most are one-shot approaches. 
The advent of agents and agentic workflows opens the door to interactive vibe modeling approaches like the one discussed here. 
But we still need to learn how to best leverage agentic capabilities and collaborate with one (or more) agent(s)to infer better models. 
Aspects like:
\begin{itemize}
\item{What types of conversations and questions the agents should have with the domain experts to validate the model being inferred? Or to disambiguate and complete the natural language description?}
\item {How to train agents for the modeling domain? What type of Reinforcement learning strategies could be useful to create specialized modeling agents?}
\item {What datasets should be created and provided to the LLMs used by the agents to improve their training (e.g. extending \cite{LopezIC22})? Or to have fine-tuned LLMs with a better understanding of the modeling concepts and tasks?}
\item {How these agents can effectively collaborate on partial models to complement and improve their own suggestions? How many agents should be involved, depending on the complexity of the model to be inferred?}
\item {How to evaluate the quality of the models inferred by the agents? And how to use that information to choose the best modeling agent for the task at hand?}
\end{itemize} 
still need to be addressed.

\subsection{Vibe modeling of other types of models, especially AI models}
Most of the current works evaluate the performance of LLMs on specific types of models, mainly class diagrams and workflow models. 
But it is unclear how good are LLMs (and the agents built on top of them) when it comes to inferring other types of diagrams, such as (UML) collaboration diagrams, architectural diagrams, user interface models,\dots.
These types of models are less popular and therefore less present in the data the LLMs have been trained on.
We need more research to understand the limitations of LLMS when it comes to this kind of models and how the aspects mentioned in the previous point would need to be adapted to cover these types of diagrams.

The situation is even worse when it comes to the modeling of the AI components integrated in a software system, also known as a smart software system \cite{lowcodesmartsoftware}.
These systems require new types of models. For instance, chatbot models \cite{planas21,sara24} for chatbot interfaces, biases requirements for the AI components \cite{langbite,disipio24}, modeling of neural networks \cite{DaoudiAC25}, etc.
The lack of standards for this type of models is an additional challenge when teaching agents how to infer them.

\subsection{Uncertainty and traceability}
Uncertainty modeling \cite{TroyaMBV21} should be considered a first-level concern. 
Indeed, when agents and LLMs are part of the modeling process, all model proposals come with a certain level of uncertainty. This confidence score should be stored together with the element. 
And for the same reason, we must be able to explain where that number came from. We need to keep full traceability of the model evolution. We should be able to explain who proposed and approved each model change. 
Similar to what was proposed in Collaboro \cite{izquierdo2016collaboro} but applied at the model level, you could link to the model elements the list of change proposals for that element, the user (human or agent) behind the change 
and the full list of users that voted in favor of the change when we are in a collaborative scenario. 

\subsection{Adapting vibe modeling to different user profiles}
Vibe modeling could be used by different types of users, from domain experts, with limited technical expertise, to software engineers with deep modeling expertise but limited domain knowledge for the domain targeted by the system-to-be.
Each profile may prefer a different type of interaction with the modeling agent/s. In the former, the agent should be able to explain the model in a way that is easy to understand for the domain expert. 
In the latter, the agent should offer a more direct approach where the user can directly validate model excerpts and where the agent may be more useful as a domain expert answering domain questions from the modeler based on its internal knowledge.

This is similar to the no-code/low-code discussion, where we also have these two types of profiles, and platforms end up offering a combination of both as they are not mutually exclusive. 
For instance, as depicted in our vibe modeling approach, we could have a non-technical user interacting with the agent with the occasional participation of a modeling expert for more complex modeling decisions, depending on the criticality of the domain.

\subsection{Native integration of agents in low-code platforms}
To provide a more natural interaction flow, we advocate for the integration of MCP clients in low-code platforms. This would enable the modeler to seamlessly switch between a "traditional" modeling flow and a vibe modeling one. 
Having a chatbot widget in the modeling UI would facilitate the adoption of vibe modeling as the user would not feel a disruption when interacting with the embedded agent. 

At the moment, very few low-code platforms offer some type of AI assisted modeling support
\footnote{Two exceptions are Mendix, which has recently introduced Maia \url{https://www.mendix.com/platform/ai/aiad/}, and OutSystems with Mentor \url{https://www.outsystems.com/low-code-platform/mentor-ai-app-generation/}}, 
even if we can foresee that more and more tools will be quickly adding this type of capability.

Nevertheless, we expect most of these vendors to release proprietary solutions, with agents directly integrated with the internal tool vendor APIs and focused on the specific types of modeling languages provided by the tool. 
We hope the popularization of MCP servers, as the one proposed herein, will help to prevent this situation, at least for open source solutions that may be more open to favor modeling interoperability. 

Note that, in our vision, the widget embedded in the tool should interact with the MCP server and not directly with the tool itself. We aim for a collaborative scenario where the agent acts as yet another user collaborating with the modelers in real-time. 
This offers more flexibility and portability as the same agent can be used from within the low-code platform but also from inside other tools (e.g. Cursor or any other IDE) that support the MCP protocol, combining the best of both worlds.

\section{Conclusions and further work}
\label{section:conclusions-and-future-work}

This paper has introduced the concept of vibe modeling and how it can enable a new style of low-code software development combining the benefits of AI and model-driven techniques. 
While this new development approach has still many shortcomings, we believe it shows promise and could contribute to reestablish the importance of (conceptual) modeling in front of current trends favoring direct "vibe coding" of the applications with all the risks this implies for the quality of the final system. 

As further work, we plan to address the roadmap outlined above to facilitate the adoption of vibe modeling and continue refining these ideas based on the feedback of the vibe modelers.


\bibliographystyle{splncs04}
\bibliography{refs}

\end{document}